\documentclass[10pt,twocolumn,twoside]{IEEEtran}
\usepackage{amsfonts}
\usepackage{amsmath}
\usepackage{amssymb}
\usepackage{epstopdf}
\usepackage{color, colortbl, multirow}
\definecolor{Gray1}{gray}{0.9}

\usepackage{subfigure}

\usepackage[dvips]{graphicx} 
\usepackage{pdflscape, algorithm, algpseudocode, cite, booktabs}

\ifCLASSINFOpdf

\else

\fi

\begin{document}
\title{Identification of Human Breathing-States Using Cardiac-Vibrational Signal for m-Health Applications}
\author{Tilendra Choudhary,~\IEEEmembership{Member,~IEEE}, L.N. Sharma,
       M.K. Bhuyan,~\IEEEmembership{Senior Member,~IEEE}, and Kangkana Bora

\thanks{T. Choudhary, L.N. Sharma, and M.K. Bhuyan are with the Department of Electronics and Electrical Engineering, Indian Institute of Technology Guwahati, India-781039 (e-mails: \{tilendra, lns, mkb\}@iitg.ac.in).

K. Bora is with the Department of Computer Science and Information Technology,
Cotton University, Guwahati, India-7810001 (e-mail: kangkana.bora@cottonuniversity.ac.in)} }

\markboth{}
{Choudhary \MakeLowercase{\textit{et al.}}: Identification of Human Breathing-States Using Cardiac-Vibrational Signal for m-Health Applications}

\maketitle

\begin{abstract}
In this work, a seismocardiogram (SCG) based breathing-state measuring method is proposed for m-health applications. 
The aim of the proposed framework is to assess the human respiratory system by identifying degree-of-breathings, such as breathlessness, normal breathing, and long and labored breathing.
For this, it is needed to measure cardiac-induced chest-wall vibrations, reflected in the SCG signal.
Orthogonal subspace projection is employed to extract the SCG cycles with the help of a concurrent ECG signal. 
Subsequently, fifteen statistically significant morphological-features are extracted from each of the SCG cycles. These features can efficiently characterize physiological changes due to varying respiratory-rates. 
Stacked autoencoder (SAE) based architecture is employed for the identification of different respiratory-effort levels.
The performance of the proposed method is evaluated and compared with other standard classifiers for 1147 analyzed SCG-beats.
The proposed method gives an overall average accuracy of 91.45\% in recognizing three different breathing states.
 The quantitative analysis of the performance results clearly shows the effectiveness of the proposed framework.
It may be employed in various healthcare applications, such as pre-screening medical sensors and IoT based remote health-monitoring systems.
\end{abstract}


\section{Introduction}
\noindent
\IEEEPARstart{T}{he} development of alarming devices for health monitoring via body area networks (BANs) has been receiving substantial interest recently.
As an m-health application, the automatic breathing-state assessment system can be employed in many electronic devices like tablets, smart phones, e-whiteboards, smart watches, and health-bands \cite{bai2012}.
Recent cardiovascular studies suggest that seismocardiography (SCG) has greater potential to be a diagnostic tool for early prediction of cardiac diseases in wearable healthcare \cite{rienzo2013_02, taebi2019}. 
The SCG signal measures cardiac mechanical events by recording cardiac-induced chest-wall vibrations \cite{inan2015, sorensen2018nature}.
These cardiovascular events are opening and closure of heart valves, blood filling and ejection through heart-chambers, and so on. 
The SCG is found somewhat advantageous over earlier cardiac modalities such as electrocardiography (ECG) and phonocardiography (PCG) \cite{choudhary2018JBHI}.
The wearable healthcare appliances integrated with an SCG-based system have a great potential for wireless wearable BANs \cite{rienzo2013_02}.  
Many factors affect the SCG signal morphology, such as body movement, posture, and respiration.
This study mainly focuses towards the analysis of SCG morphology under varying respiratory-effort levels.

Long term irregularities in respiratory rhythms often affect the heart and the lung functions.
Hence, identification of breathing patterns is an essential task to avoid the related diseases. 
In medical diagnosis, breathing pattern and heart-rate are considered as primary screening tools, which provide symptoms of various life-threatening diseases, including cardiovascular diseases like arrhythmias, cardiac arrest and sepsis, and diseases due to lung dysfunctions such as asthma, pneumonia, chronic obstructive pulmonary disease (COPD), hypercarbia and pulmonary embolism \cite{charlton2018}.
Fear, anxiety and extensive exercises could also produce abnormal breathing symptoms even in healthy individuals.
During the breathing inhalation time, the diaphragm contracts and moves down, the chest surface expands, pressure in the intrathoracic cavity reduces, each of the lungs inflates, and the heart moves almost linearly with the displacement of diaphragm \cite{taebiThesis}.
The hemodynamic variations, such as changes in blood volume, turbulence and pressure caused due to decreased intrathoracic pressure affect the morphological structure of the SCG signal \cite{taebiThesis}.
The morphology of an SCG signal is affected by different respiratory conditions, and so, the SCG can be used not only to measure cardiac health, but also to assess lung fitness  
\cite{taebi2019}, \cite{choudhary2019TENSYMP}.
In our previous work \cite{choudhary2019TENSYMP}, significant changes in SCG morphologies are shown for two different respiratory conditions. 
{Breathing patterns may include various states, such as normal breathing, breathlessness, long and labored breathing, and other irregular breathing rhythms.} 
The breathlessness and long labored breathing are abnormal breathing patterns, which are often observed in acute/chronic dyspnea and orthopnea cases \cite{DyspneaWeb1}.
In the current scenarios, COVID-19 also exhibits abnormal breathing symptoms with its progression.
The severe abnormalities in lung and heart are the major causes of these conditions in most of the cases. 
It is to be mentioned that the physical examinations cannot always diagnose these conditions \cite{DyspneaWeb1}.
The SCG would be helpful in establishing physiological-relationship for cardiorespiratory system. This may also be applicable to cardiopulmonary exercise testing (CPET).

\begin{figure*}
\centering
\includegraphics[width=18cm]{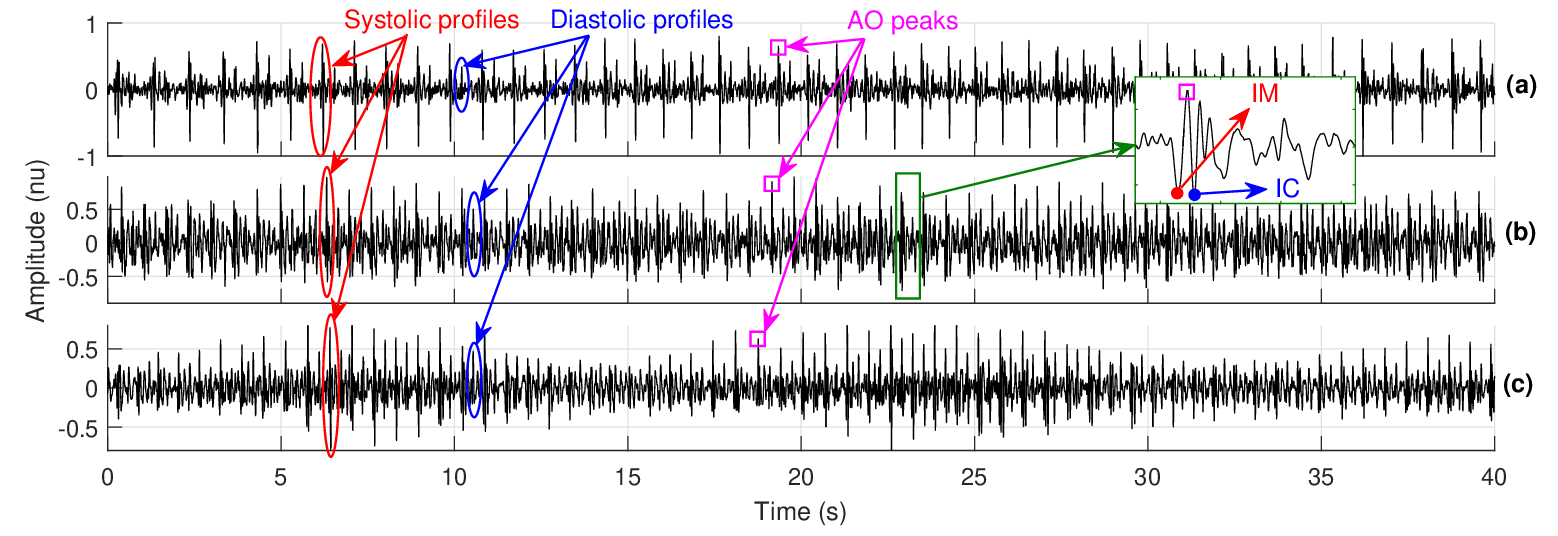}
\vspace{-0.3cm}
\caption{Annotated SCG signals in three breathing scenarios: (a) breathlessness, (b) normal breathing, and (c) long breathing. Morphological variations are observed in all the three respiratory patterns, where signals were collected from a single subject.}
\label{fig:Signals_SBNBLB}
\vspace{-0.1cm}
\end{figure*}

In the existing literature, a few research works have been suggested using ECG or SCG signals for detection of respiratory information, such as extraction of breathing rate sequence and detection of sleep apnea.
To identify the respiratory inhale and exhale phases, the SCG signal may also be used \cite{zakeri2017, pandia2013, zakeri2015, alamdari2016, alamdari2015}.
Zakeri {\it et al.} devised an approach to analyze the SCG beats for the recognition of respiratory phases \cite{zakeri2017}. 
In this approach, an SCG beat is segmented into identical sized blocks in temporal and spectral domains, and average values from the blocks are used as features. Thereafter, support vector machine (SVM) is used to select an optimal feature-group which gives a higher identification-rate.
Another method was proposed in this direction, which considers an averaged value of 512 data points of each of the systolic-profiles as a feature, and subsequently, an SVM is used for identification of breathing phases \cite{zakeri2015}.
The aforementioned schemes use R-peaks of temporally concurrent ECG signals for the segmentation of SCG cycles.
In \cite{alamdari2015}, it was demonstrated that all SCG cycles can be categorized into inhale and exhale phases with the help of a respiratory signal. 
In this direction, a frequency-domain SCG signal analysis is suggested by Pandia {\it et al.} \cite{pandia2013}.
The entire frequency range is splitted into two spectral bins corresponding to 5 and 10 Hz, and discrimination between the inhalation and exhalation phases is statistically done in the 10--40 Hz frequency range.
As a preliminary work, we presented a method for characterization of two breathing states by analyzing morphological differences of an SCG waveform \cite{choudhary2019TENSYMP}. 
However, the variation of morphological characteristics of the SCG signal due to different respiratory conditions is still needed to be extensively investigated.

 The objective of this study is to propose an SCG-based breathing-state detector for m-healthcare applications. 
The proposed framework is designed to identify different  breathing patterns, namely breathlessness, regular/normal respiration, and long and labored breathing.
The labored breathing is an abnormal pattern characterized by a symptom of increased breathing effort. 
All these patterns are abbreviated as SB, NB, and LB for stopped, normal and long breathing, respectively. 
The proposed breathing-state detector needs a concurrent ECG signal to extract the SCG cycles. 
A set of statistical-, amplitude-, time-, and spectral-based features of the SCG signal is extracted.
In our method, stacked autoencoder (SAE) based neural network (NN) architecture is used for identification of different breathing levels.
The rest of the paper is organized as follows:
Section II presents the proposed methodology. The
experimental results are presented in Section III. Finally, conclusions are drawn in Section IV.

\section{Proposed Breathing State Detection Method}
\begin{figure*}
\includegraphics[height=8.9cm, width=2.9cm]{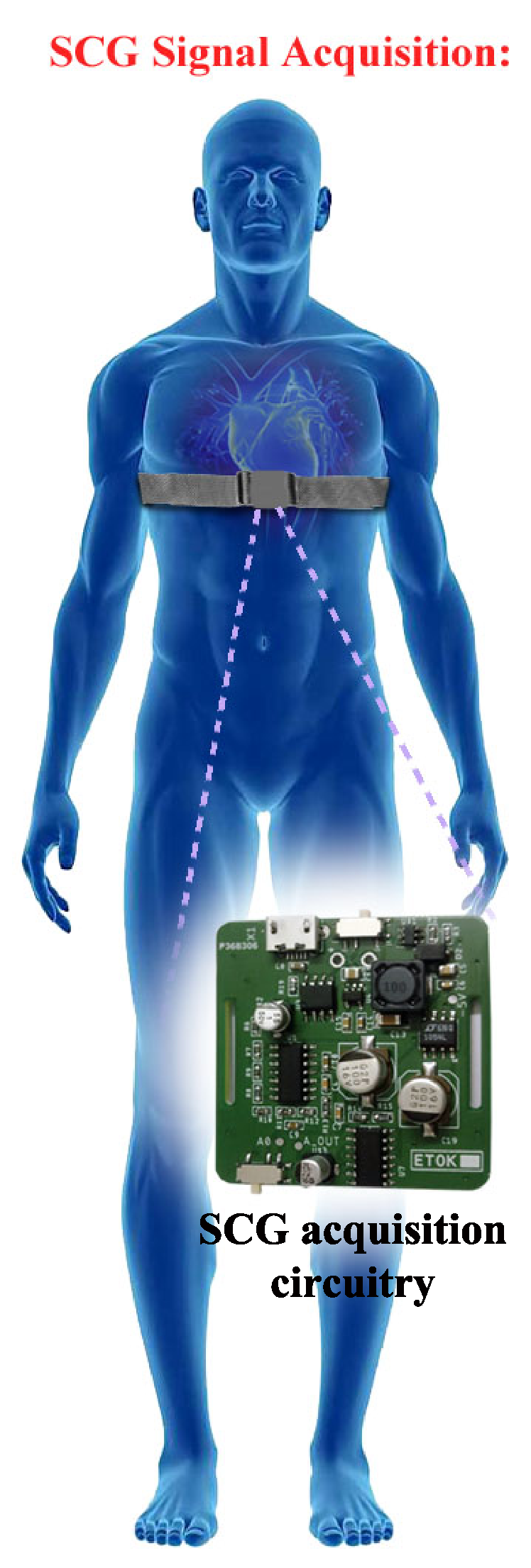}
~~\includegraphics[width=14cm]{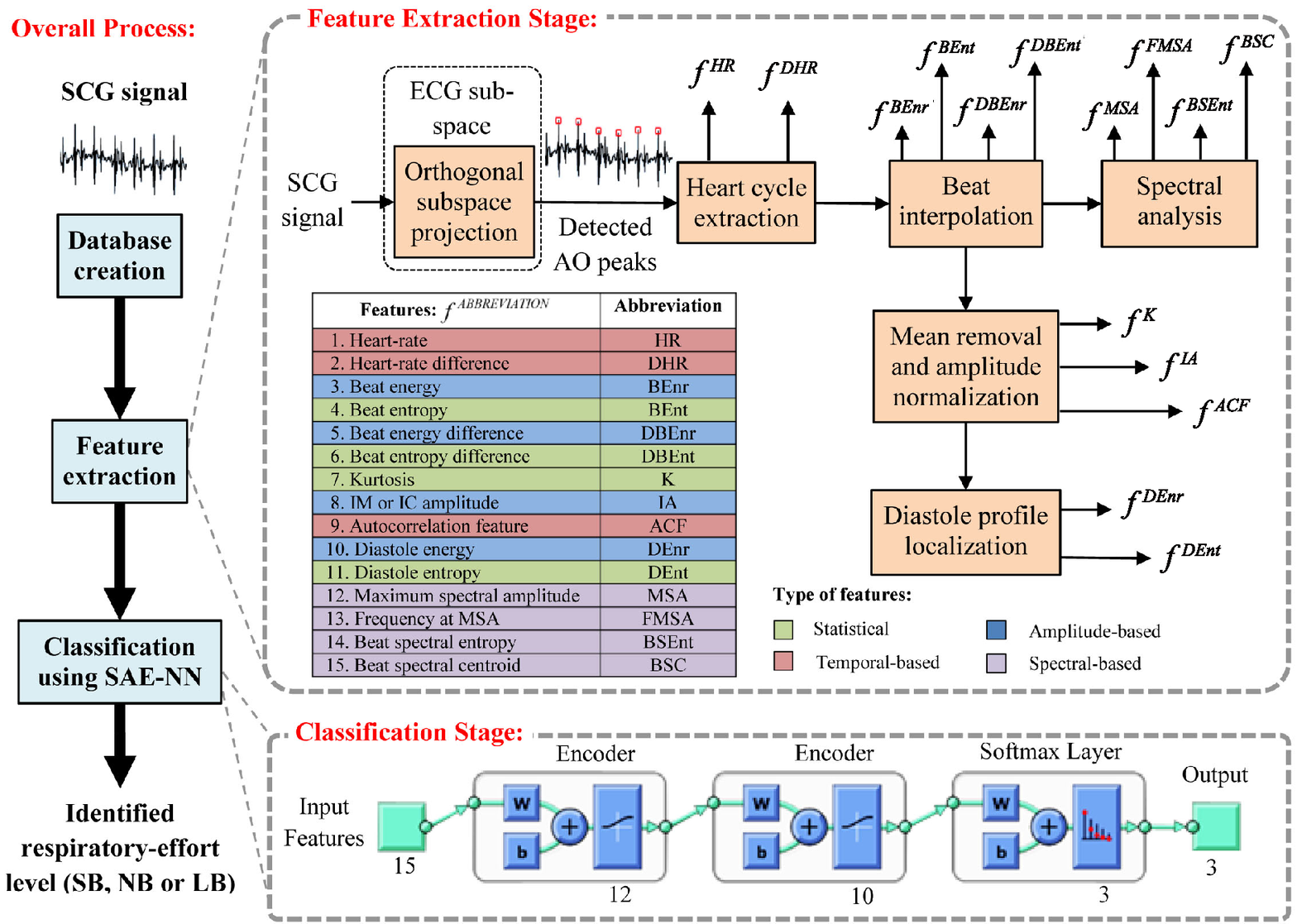}
\caption{Overview of the proposed method for identification of respiratory-effort levels. The numeric details shown in the classification stage represent numbers of nodes at different layers of SAE-NN classifier. For instance, there are 15 input features, and both sequential encoders and softmax layer produce 12, 10 and 3 output nodes, respectively.}
\label{fig:overview}
\end{figure*}

The SCG beat morphologies can indicate respiratory-effort levels. As shown in Fig.~\ref{fig:Signals_SBNBLB}, the waveform characteristics of SCG signals changes for SB, NB, and LB breathing conditions.
More specifically, SCG signals in breathlessness condition have peaky-distributed beat patterns having almost constant amplitudes and regular heart-rhythms, while relatively more variations of amplitude and heart-rhythm are observed during normal breathing. 
Large amplitude-modulated type beat patterns with varying heart-rates are observed in long breathing conditions.
In order to identify different breathing states, the features which can identify and segregate morphological variations of an SCG signal due to different breathing conditions need to be extracted. 
The overview of the proposed methodology is illustrated in Fig.~\ref{fig:overview}. The proposed work is carried out in three major phases. In Phase-I, signal-database was generated followed by feature extraction in Phase-II. Finally, classification is done to identify the degree of breathing levels (SB, NB and LB).

\subsection{Database Creation}
For the breathing level identification purpose, the dorso-ventral SCG and concurrent ECG (Lead-II) signals are acquired from healthy male subjects, lying in a supine position, at Electro-Medical and Speech Technology Laboratory (EMST Lab), Indian Institute of Technology (IIT) Guwahati, India.  
{Eight subjects are chosen for this study and they have following demographics}, age: 28.75$\pm$2.31~yrs, weight: 71.63$\pm$7.85~kg, height: 5'7.6"$\pm$2.6", heart-rate: 79.18$\pm$10.93~bpm.
The signals are recorded in three sessions: normal
breathing for 5~minutes (NB), holding breath for 50~s (SB), and long respiration for 2~minutes (LB). 
{However, the same duration of 40 s data from each recorded signal is taken for the study.}
So, two breathing conditions, namely breathlessness as well as long- and labored-respiratory data are artificially generated. 
The signals are sampled at a frequency of 1 kHz.
All the signals were recorded using our self built data acquisition system (DAS).   
The description of the designed DAS is provided in \cite{choudhary2019ISJ}.
The recording process was approved by the institutional ethical review board.

\subsection{Feature Extraction}
The SCG exhibits more morphological variations for different respiratory activities as compared to an ECG signal \cite{choudhary2019TENSYMP}. Hence, the SCG signal is selected for feature extraction. A number of features are extracted from an SCG signal, and these features are mainly based on statistical, amplitude, temporal and spectral information of the signal. These features can uniquely relate the SCG morphology with the respiration rate. 
The extraction process of all these features are illustrated in Fig. \ref{fig:overview}. A detailed description of feature extraction steps is provided below. 

\vspace{0.2cm}
\subsubsection{Orthogonal Subspace Projection}
The proposed method is mainly based on the extraction of SCG cycles, which relies on accurate detection of prominent AO peaks in the SCG. The AO peaks in the SCG signal correspond aortic valve opening instants of the heart. The estimation of AO peaks is performed using an orthogonal subspace projection (OSP) scheme \cite{choudhary2019BSPC}. 
{The information of SCG components that is linearly associated with its concurrent ECG can be estimated by projecting the SCG signal onto the ECG subspace.}
The ECG subspace is created by the original ECG signal and its delayed versions \cite{choudhary2019BSPC}. 
\begin{figure}
\centering
\includegraphics[width=8.8cm]{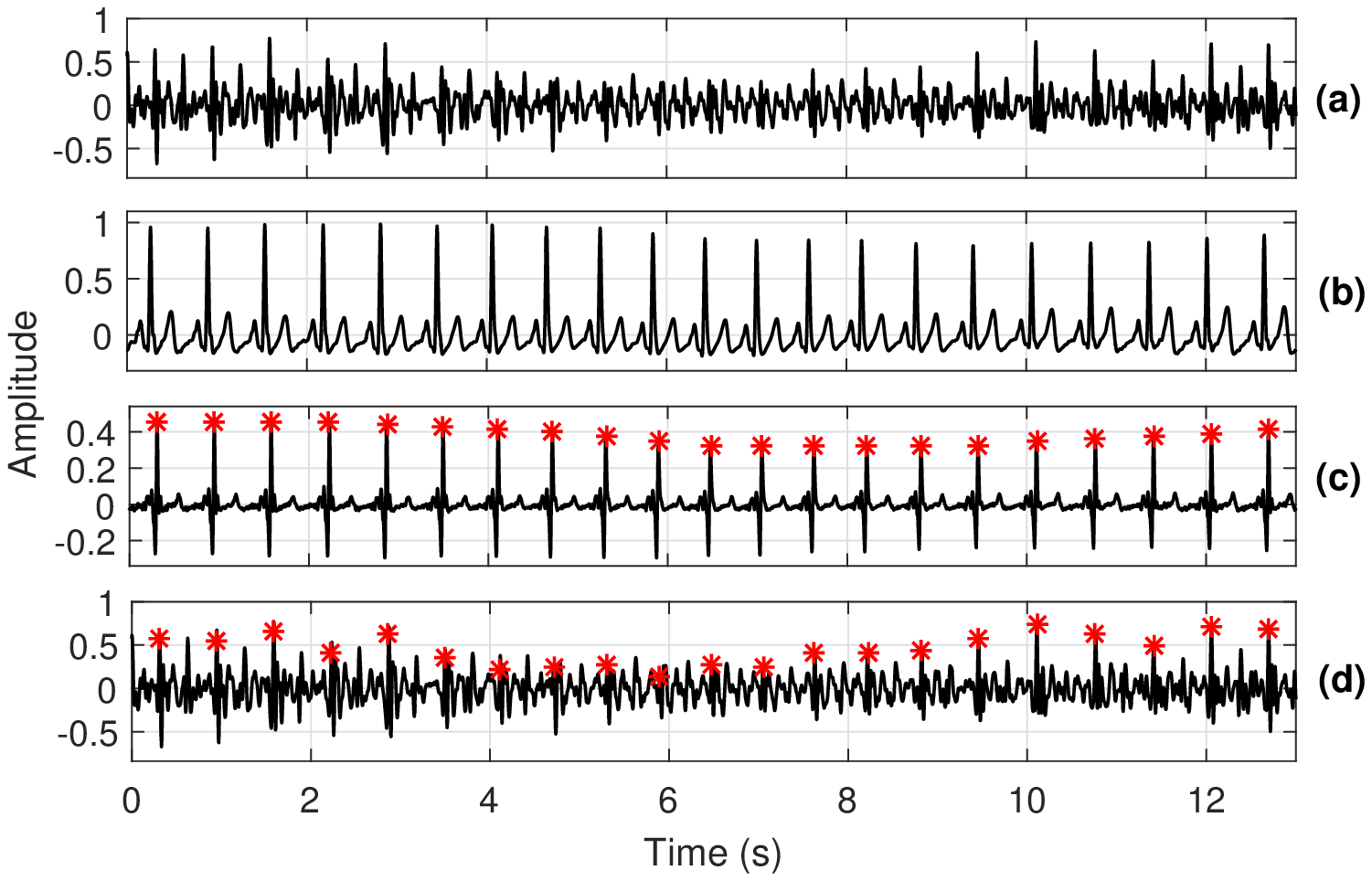}
\vspace{-1.6cm}
\caption{AO peak detection process using orthogonal subspace projection. (a) an SCG signal $\textbf{s}$, (b) concurrent ECG signal for creation of its subspace $\mathbb{U}$, (c) projected sequence $\widehat{\textbf{s}}$ and its estimated peaks using the FOGD scheme, and (d) detected AO peaks in the SCG.}
\label{fig:AOdetection}
\vspace{-0.2cm}
\end{figure} 
The linear relationship of an SCG signal $\textbf{s}$ and the corresponding ECG subspace $\mathbb{U}$ can be expressed as \cite{choudhary2019BSPC}:
\begin{equation}
\mathbb{U}\textbf{x}=\textbf{s}
\end{equation}
\noindent
The best estimate of $\textbf{s}$ on subspace $\mathbb{U}$ is found by using a subspace projection as \cite{choudhary2019BSPC}:
\begin{equation}
\widehat{\textbf{s}} = \mathbb{U}  (\mathbb{U}^{T} \mathbb{U})^{-1} \mathbb{U}^{T}\textbf{s}
\end{equation}
\noindent
Finally, a first order Gaussian differentiator (FOGD) based logic is applied to the projected sequence $\widehat{\textbf{s}}$, which  ultimately indicates the locations of AO peaks in the SCG signal. The entire AO peak detection process is shown in Fig. \ref{fig:AOdetection}. 

\vspace{0.2cm}
\subsubsection{Heart Cycle Extraction}
In our proposed method, the estimated AO instants are used to extract heart cycles ($\text{HC}$) in the SCG signal. 
Changes of breathing levels indirectly alter oxygen content present in the blood, and so, the heart pumping rates change accordingly. 
Thus, heart rate (HR) variations can indicate respiratory effort levels.
In order to extract the heart cycles, the intervals between consecutive AO instants are estimated and two different features, namely heart rate ($ f^{HR} $) and heart rate difference ($ f^{DHR} $) are estimated as follows:
\begin{equation}
\text{HC}_i= \text{AO}_{i+1} - \text{AO}_{i}
\end{equation}
\begin{equation}
f^{HR}_i  = 60 / \text{HC}_i
\end{equation}
\begin{equation}
f^{DHR}_i = \big|f^{\overline{HR}}_{i} - f^{\overline{HR}}_{i-\text{D}}\big|
\end{equation}
\noindent
where, $|\cdot|$ is absolute value operator; $i=1,2,...,(\text{P}-1)$, and $\text{P}$ denotes total number of AO peaks present in the SCG segment. $f^{\overline{HR}}$ makes $f^{HR}$ to be a circular sequence, and the delay parameter $\text{D}$ is judiciously chosen as 3 for identifying long breathing patterns.

\vspace{0.2cm}
\subsubsection{Beat Interpolation}
After utilizing temporal information of an SCG signal for indicating heart beats,
 all the beat-durations are normalized to a fixed length. 
The resulting interpolated SCG beats are used to estimate the following features: beat energy ($f^{BEnr}$), beat entropy ($f^{BEnt}$), beat energy difference ($f^{DBEnr}$), and beat entropy difference ($f^{DBEnt}$).
The beat energy is expressed as:
\begin{equation}
f^{BEnr}_i=\dfrac{1}{L}\sum_{l=0}^{L-1} \left( \bar{\textbf{s}}^i[l] \right)^2
\end{equation}
\noindent
where, $\bar{\textbf{s}}^{i}[l]~ (l=0,1,..,L-1)$ denotes $i^{th}$ interpolated SCG beat, $i\in [1,\text{P}-1]$.
For the computation of $f^{DBEnt}$ feature, the amplitude level of beat $\bar{\textbf{s}}[l]$ is normalized, say $\bar{\textbf{s}}_1[l]$, such that $\bar{\textbf{s}}_1[l] \in [0,1]$ and  $\sum_l\bar{\textbf{s}}_1[l]=1$. It expresses the distribution of relative components in a beat as:
\begin{equation}
\bar{\textbf{s}}_1[l] \longleftarrow \dfrac{\bar{\textbf{s}}[l]-\text{M}_{\bar{\textbf{s}}}}{ \sum_{l=0}^{L-1} \left(  \bar{\textbf{s}}[l]-\text{M}_{\bar{\textbf{s}}}  \right)}
\end{equation} 
\noindent
where, $\text{M}_{\bar{\textbf{s}}}=\displaystyle\mathop{min}_{l=0:L-1}(\bar{\textbf{s}}[l])$. Hence, the beat entropy can be expressed as:
\begin{equation}
f^{BEnt}_i=-\sum_{l}\bar{\textbf{s}}^i_1[l] log(\bar{\textbf{s}}_1^i[l])
\end{equation}
\noindent
The difference features corresponding to beat energy and beat entropy are computed as:
\begin{equation}
f^{DBEnr}_i=\big|f^{\overline{BEnr}}_{i} - f^{\overline{BEnr}}_{i-\text{D}}\big|
\end{equation}
\begin{equation}
f^{DBEnt}_i=\big|f^{\overline{BEnt}}_{i} - f^{\overline{BEnt}}_{i-\text{D}}\big|
\end{equation}
\noindent
where, $f^{\overline{BEnr}}$ and $f^{\overline{BEnt}}$ denote circular versions of $f^{BEnr}$ and $f^{BEnt}$ sequences, respectively. 

\vspace{0.2cm}
\subsubsection{Mean Removal and Amplitude Normalization}
After the estimation of beat energy and entropy features, DC-offset is subtracted from $\bar{\textbf{s}}_1[l]$, and the amplitude is normalized by its maximum value. This normalized beat, also denoted by $\bar{\textbf{s}}_2[l]$, is shown in Fig. \ref{fig:beat_normalized}. Under this, three morphological features, namely kurtosis ($ f^{K} $), IM or IC amplitude ($ f^{IA} $) and auto-correlation feature ($ f^{ACF} $) are extracted from the SCG beat $\bar{\textbf{s}}_2[l]$, which are described as follows:

\begin{figure}
\includegraphics[width=8.8cm]{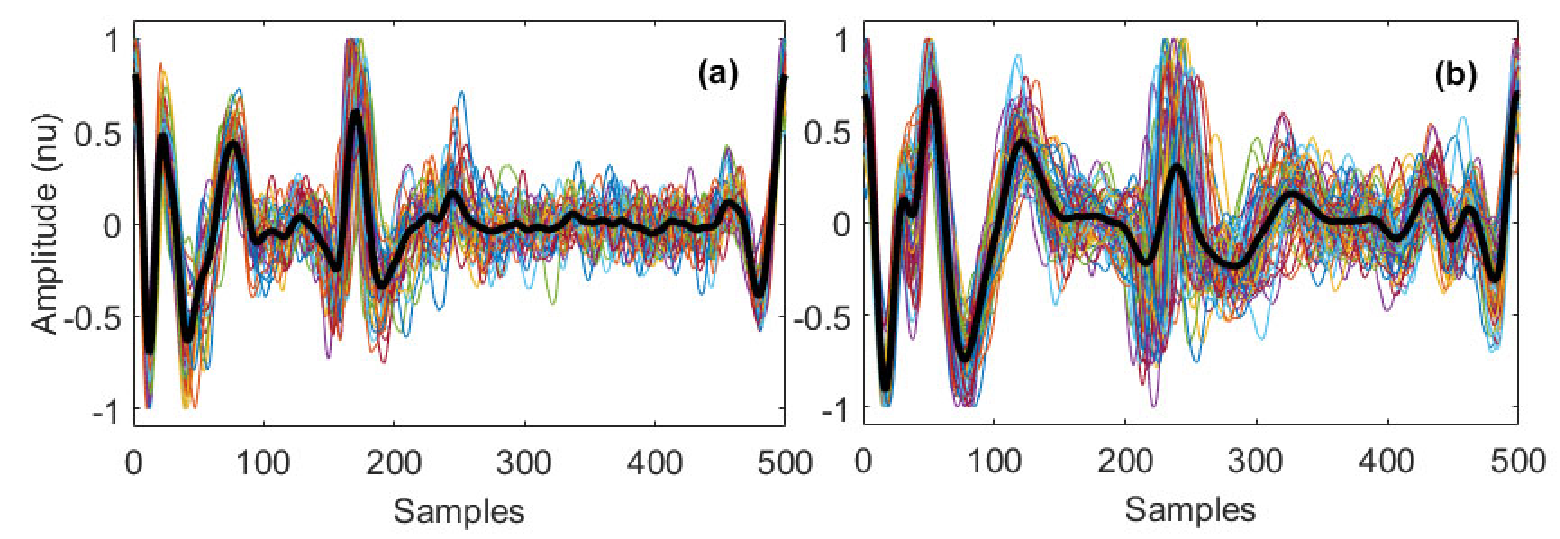}
\caption{Interpolated SCG beats after DC-offset removal and amplitude normalization for two subjects in (a) and (b). All the inter-beats are displayed along with their ensemble-averaged beat (in dark black colour).}
\label{fig:beat_normalized}
\vspace{-0.2cm}
\end{figure}

\begin{itemize}
\vspace{0.1cm}
\item \textbf{Kurtosis (K):} It estimates the peakedness of the distribution for SCG beat $\bar{\textbf{s}}_2[l]$. It is defined as \cite{choudhary2018JBHI}:
\begin{equation}
f^{K}_i = \dfrac{\dfrac{1}{L} \displaystyle\sum_{l=0}^{L-1} \left(\bar{\textbf{s}}_2^i[l] - \text{M}_{\bar{\textbf{s}}_2^i}\right)^{4}}{\left(\dfrac{1}{L} \displaystyle\sum_{l=0}^{L-1} \left(\bar{\textbf{s}}_2^i[l] - \text{M}_{\bar{\textbf{s}}_2^i}\right)^{2}\right)^{2}}
\end{equation}

\noindent
where, $\text{M}_{\bar{\textbf{s}}_2^i} = \frac{1}{L} \sum_{l=0}^{L-1} \bar{\textbf{s}}_2^i[l]$. As compared to other breathing patterns, larger Kurtosis values correspond to breathlessness conditions \cite{choudhary2019TENSYMP}.

\vspace{0.1cm}
\item \textbf{Autocorrelation feature (ACF):} The ACF feature is extracted from each of the beats as follows \cite{choudhary2018JBHI}: 
\begin{equation}
f^{ACF}_i=\sum_{l=-\infty}^{\infty}\{\bar{\textbf{s}}_2^i[l]\bar{\textbf{s}}_2^i[l+P]\}\label{Eq:ACC2}
\end{equation}
\noindent
where, the parameter $P$ denotes a fixed lag. 
The ACF feature can detect interbeat variabilities and body vibrations resulted from varying respiration rates.

\vspace{0.1cm}
\item \textbf{IM/IC amplitude (IA):} The IA is extracted by computing the maximum negative signal-strength of each beat $\bar{\textbf{s}}_2$ \cite{choudhary2019TENSYMP}. It indicates the amplitude of either IM or IC fiducial point. This feature is used to measure the amplitude sharpness of an SCG cycle induced by varying breathing-pattens. It can be expressed as:
\begin{equation}
f^{IA}_i=\displaystyle\mathop{min}_{l=0:L-1}(\bar{\textbf{s}}_2^i[l])
\end{equation}

\end{itemize}

\vspace{0.2cm}
\subsubsection{Diastole Profile Localization} 
\begin{figure}
\includegraphics[width=8.8cm]{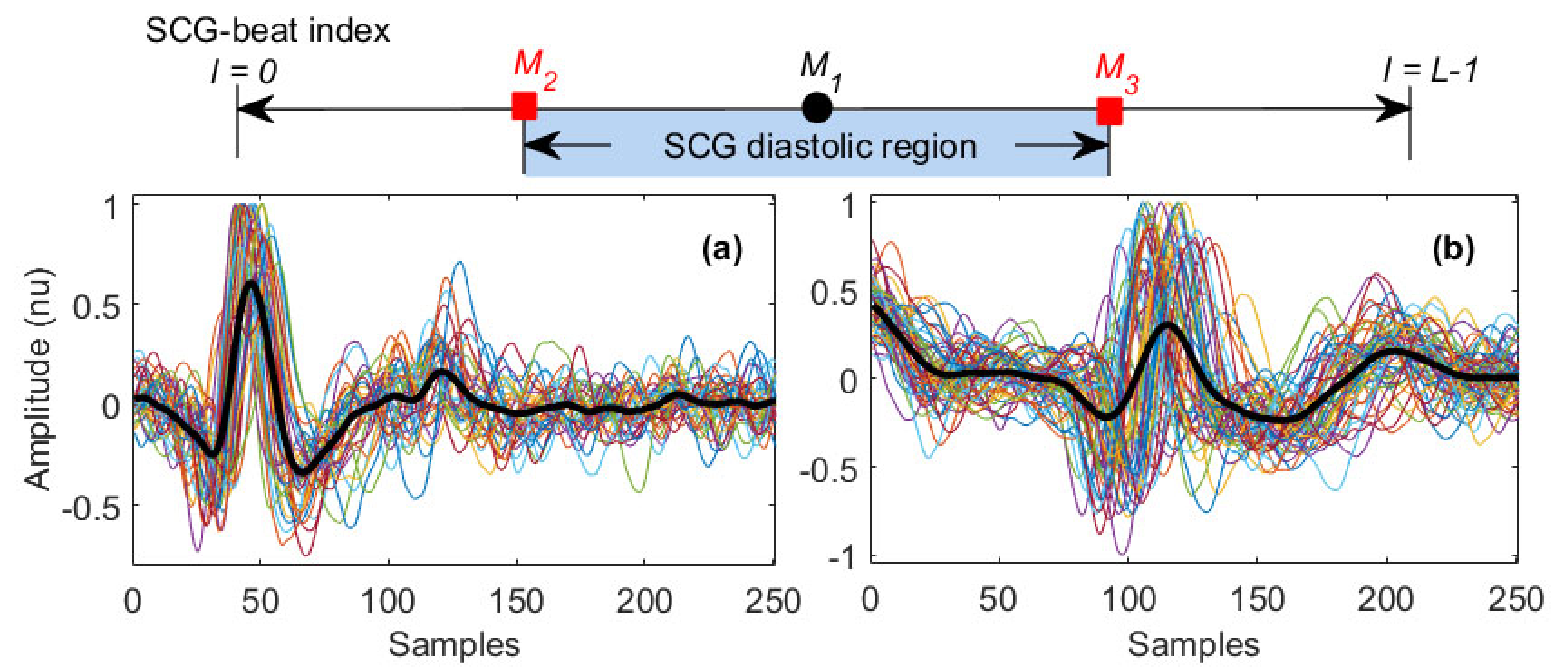}
\caption{Segmented SCG diastole-profiles after DC-offset removal and amplitude normalization for two subjects in (a) and (b). All the inter-beats are displayed along with their ensemble-averaged beat (in dark black colour).}
\label{fig:Diastole_segment}
\vspace{-0.2cm}
\end{figure}

In order to extract the diastolic features, 
the SCG diastole is segmented as follows. Initially, the beat $\bar{\textbf{s}}_2$ is divided at its middle (say $M_1$ point). The SCG systole can be localized in a segment between $M_2$ and $M_3$ points, where $M_2$ and $M_3$ are the mid-points of segments between start-point and $M_1$, and between $M_1$ and end-point, i.e., 
\begin{equation}\nonumber
M_1=\dfrac{(L-1)}{2};~~M_2=\dfrac{(L-1)}{4};~~M_3=\dfrac{3(L-1)}{2}
\end{equation}
\noindent
The segmentation of diastolic region is also shown in Fig.~\ref{fig:Diastole_segment} along with diastole profiles segmented from two subjects.
To capture morphological variabilities especially in diastole profiles, two features, diastole Energy ($ f^{DEnr} $) and diastole entropy ($ f^{DEnt} $) are extracted. 
The expression of $ f^{DEnr} $ is given as follows:
\begin{equation}
f^{DEnr}_i = \frac{1}{M_3-M_2+1} \sum_{l=M_{2}}^{M_{3}} \left(\bar{\textbf{s}}_2^i[l]\right)
^2
\end{equation}

\noindent
Usually, the SCG-diastole produces relatively smallest energy in breathlessness condition as compared to other breathing conditions. Thus, $f^{DEnr}$ can be a dominant feature to detect breathing events. 
Subsequently, diastole entropy ($ f^{DEnt} $) feature is extracted on $\textbf{d}_1[n]$ ($\textbf{d}_1[n] \in [0,1]$), which has a zero minima and it is normalized by its elemental-sum value as:
\begin{equation}
\textbf{d}[n] = \bar{\textbf{s}}_2[M_2,M_2+1,M_2+2,...,M_3]
\end{equation}
\begin{equation}
{\textbf{d}}_1[n] \longleftarrow \dfrac{{\textbf{d}}[n]-\text{M}_{{\textbf{d}}}}{ \sum_{n=M_2}^{M_3} \left(  {\textbf{d}}[n]-\text{M}_{{\textbf{d}}}  \right)}
\end{equation} 

\noindent
where, $\text{M}_{\textbf{d}}=\displaystyle\mathop{min}_{n=0:M_3-M_2}({\textbf{d}}[n])$ and the expression for $ f^{DEnt} $ is given as:
\begin{equation}
f^{DEnt}_i=-\sum_{n}{\textbf{d}}^i_1[n] log({\textbf{d}}_1^i[n])
\end{equation}

\vspace{0.2cm}
\subsubsection{Spectral Analysis} 
Four spectral features namely, maximum spectral amplitude ($ f^{MSA} $), frequency at MSA ($ f^{FMSA} $), beat spectral centroid ($ f^{BSC} $), and beat spectral entropy ($ f^{BSEnt} $) are extracted from magnitude-spectrum of each of the interpolated beats $(\bar{\textbf{s}}_1^i[l])$.
Suppose, $\bar{\textbf{s}}_1[l]$ and $\bar{\textbf{S}}_1[f]$ are Fourier transformation pairs, and $|\bar{\textbf{S}}_1[f]|$ denotes the magnitude spectrum. Then, $ f^{MSA} $ and $ f^{FMSA} $ features are expressed as:
\begin{equation}
 f^{MSA}_i = \displaystyle\mathop{max}_{f=0,1,..,F_s/2}  \left( |\bar{\textbf{S}}_1^i[f] | \right)
\end{equation}
\begin{equation}
 f^{FMSA}_i = \displaystyle\mathop{arg~max}_{f=0,1,..,F_s/2} \left(|\bar{\textbf{S}}_1^i[f]|\right)
\end{equation}

\noindent
where, $F_s$ denotes sampling frequency of the signal.
The MSA and FMSA features characterize the beat by its dominating spectral component, which changes with varying respiration-rates. Similarly, the beat spectral centroid (BSC) and the beat spectral entropy (BSEnt) measure the center-of-gravity of the beat-spectrum and spectral randomness, respectively. Their expressions are given as follows: 
\begin{equation}
f^{BSC}_i=\dfrac{\sum_f f  |\bar{\textbf{S}}_1^i[f] | }{\sum_f  |\bar{\textbf{S}}_1^i[f] | }
\end{equation}
\begin{equation}
f^{BSEnt}_i=-\sum_{f}{\textbf{S}}^i_n[f] log({\textbf{S}}_n^i[f])
\end{equation}

\noindent
where, $\textbf{S}_n[f]$ corresponds to normalized $\bar{\textbf{S}}_1[f]$ 
i.e., $\textbf{S}_n[f]=\bar{\textbf{S}}_1[f]/(\sum\bar{\textbf{S}}_1[f])$.

Finally, all the extracted features are concatenated together to create a feature vector $(\mathbb{R}^{15})$ corresponding to each of the SCG-beats.

\subsection{Stacked Autoencoder-based Model for Identification of Breathing Conditions}

\begin{figure}
\centering
\includegraphics[width=8.8cm]{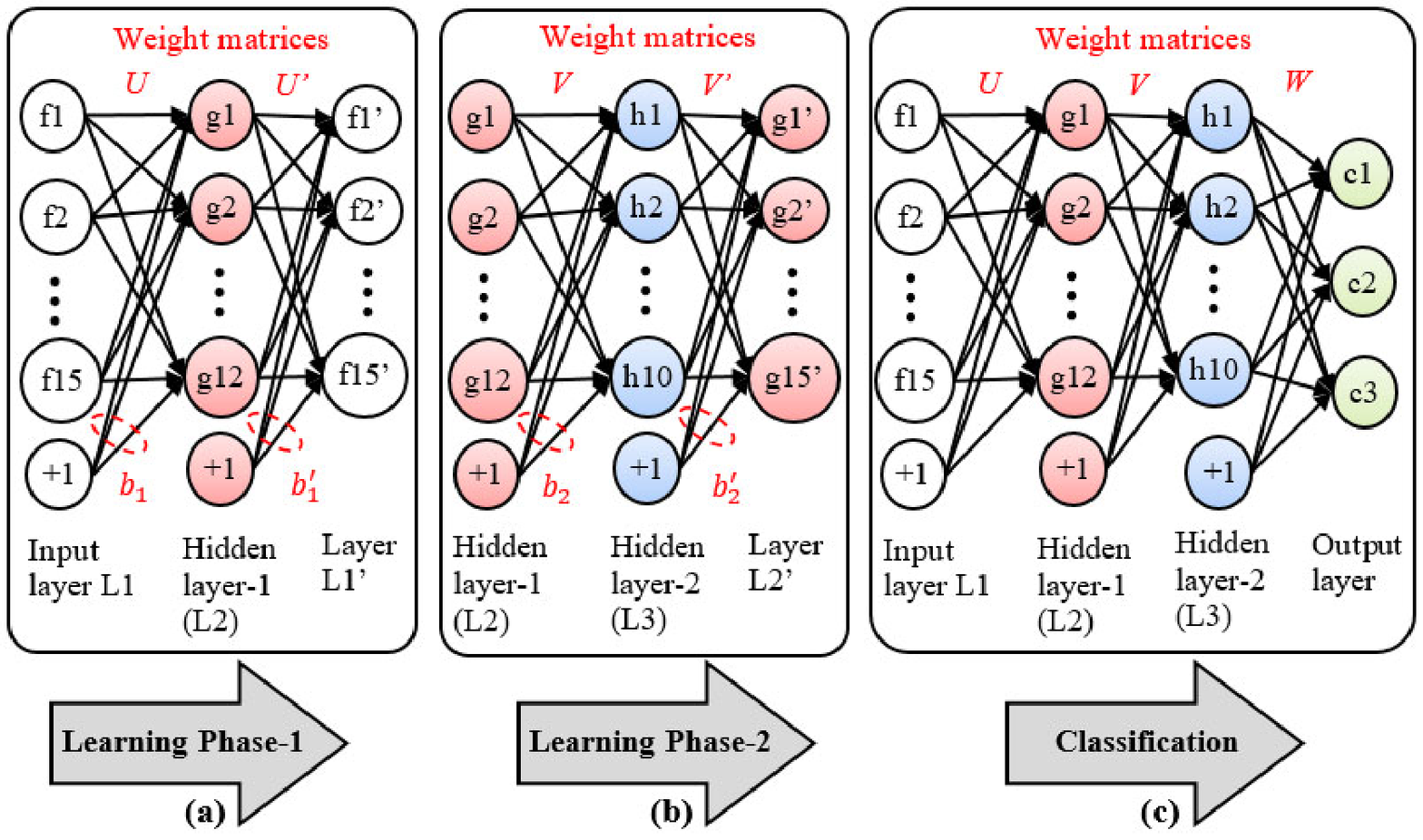}
\caption{Proposed network configuration to develop stacked autoencoder (SAE) based classifier for identification of breathing-patterns. (a) autoencoder for layer L2, (b) autoencoder for layer L3, and (c) SAE for classification.} 
\label{SAE}
\vspace{-0.2cm}
\end{figure}

Autoencoder is a neural network (NN) with an input layer, a hidden layer, and an output layer as shown in Fig. \ref{SAE}(a) and (b).
The autoencoder is trained to reconstruct input data using encoding and decoding processes \cite{ravi2017}.
Let $x$ and $b$ be the input vector and bias, respectively, then the hidden space is expressed as, $z=\phi(Ax+b)$, where `$z$' is mapping or code vector, `$A$' denotes the encoding weight matrix, and the linear or nonlinear nature of mapping can be set by activation function $\phi(\cdot)$. Similarly, the decoder maps the hidden space data to the input vector as, $\widehat{x}=\phi '(A'z+b')$.
In general, $\phi '(\cdot)$ is considered linear, since input data belongs to the space of real numbers \cite{gogna2017}. 
The learning of connection weights corresponding to encoding ($A$) and decoding ($A'$) is achieved by minimizing the cost function given in \cite{niyaz2017}:
\begin{multline}
J = \frac{1}{2R} \sum_{r=1}^{R} \sum_{k=1}^{K}  (x_{kn}-\widehat{x}_{kn})^2 +    \frac{\lambda}{2} \left( \sum_{N,K} (A)^{2} + \sum_{K,N} (A')^{2} \right. \\ \left. + \sum_{N} (b)^{2} +  \sum_{K} (b')^{2} \right) +  \beta \sum_{j=1}^{N} \text{KL}(\rho \parallel \widehat{\rho}_{j})
\label{cost}
\end{multline}

\noindent
where, input and output layers have equal $K$ nodes, and the hidden layer has $N$ nodes, thus $A\in \mathbb{R}^{N \times K}$, $A' \in \mathbb{R}^{K \times N}$, $b \in \mathbb{R}^{N \times 1}$, and $b' \in \mathbb{R}^{K \times 1}$. The used parameters $R$, $\lambda$, and $\beta$ denote total number of observations in the dataset, weight regularization, and sparsity regularization parameters, respectively. $\text{KL}$, which represents Kullback-Leibler divergence, acts as a sparsity penalty component \cite{niyaz2017}. 
The indices $\rho$ and $\widehat{\rho}$ correspond to desired and average activation values, respectively.

For classification, the sparse autoencoders are stacked together and a combined neural network architecture is created.
This is called as stacked autoencoder (SAE), where the weights are usually learned in a greedy manner \cite{gogna2017}.
Fig. \ref{SAE} shows the proposed SAE network configuration for identification of respiratory rates.
It is an ensemble of two sparse autoencoders with a soft-max classifier.
As shown in Fig \ref{SAE}, during the first learning phase, hidden space $\textsl{g}$ is trained on input feature vector $f$ to obtain the weight matrices $ U, ~U'$, and bias vectors $b_1, ~b'_1 $. While, the next hidden layer $h$ is trained on the $\textsl{g}$ space to get the weight matrices $V, ~V'$, and bias vectors $b_2, ~b'_2 $. Subsequently, the output of the last hidden layer is fed into a soft-max classifier. 
Finally, all the SAE layers are used as a single unified model, and this model is fine-tuned for performance improvement as shown in Fig. \ref{SAE}(c).

\section{Experimental Results and Discussion}
\begin{figure}
\centering
\includegraphics[width=3.8cm]{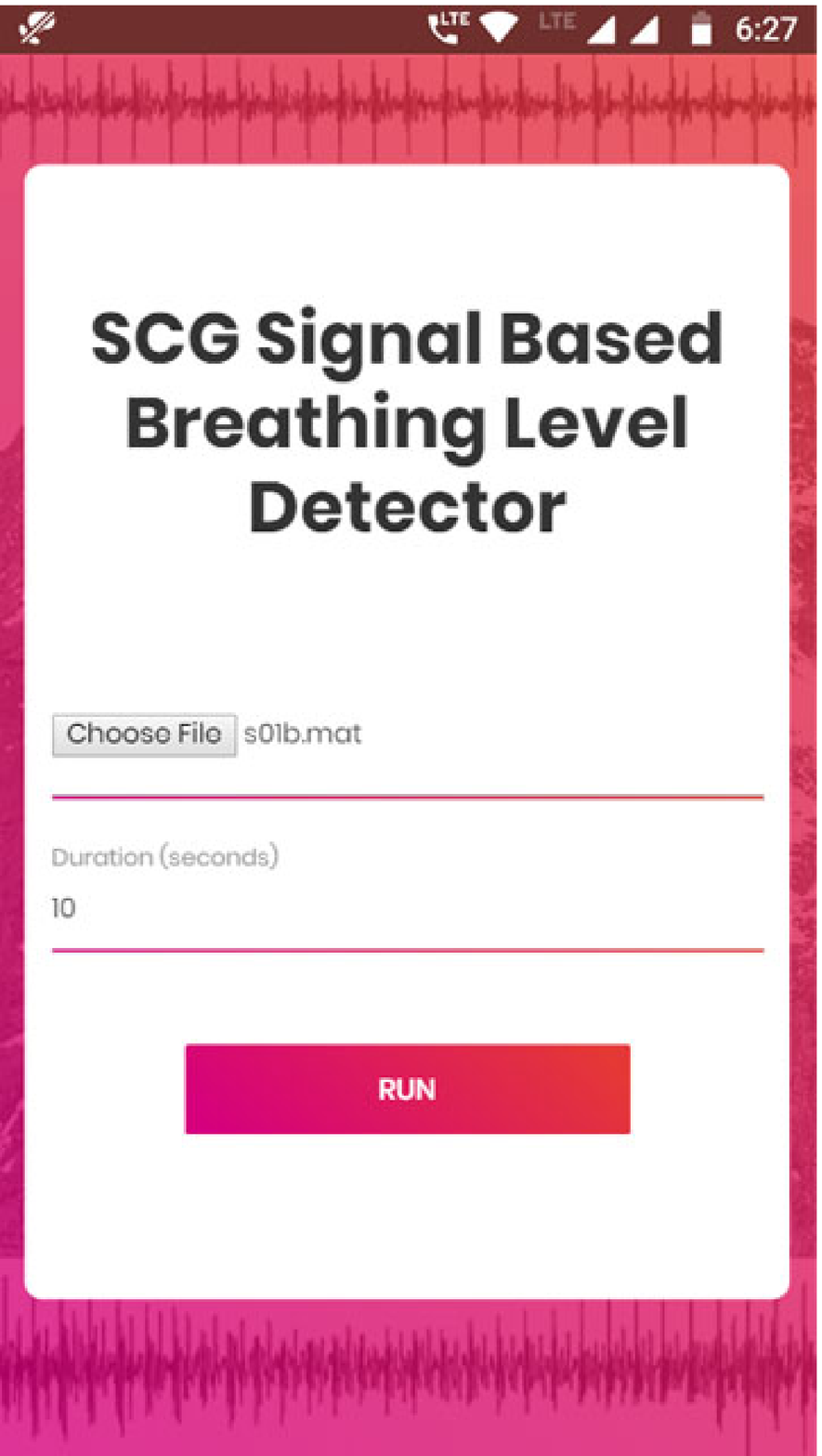}~~~
\includegraphics[width=3.8cm]{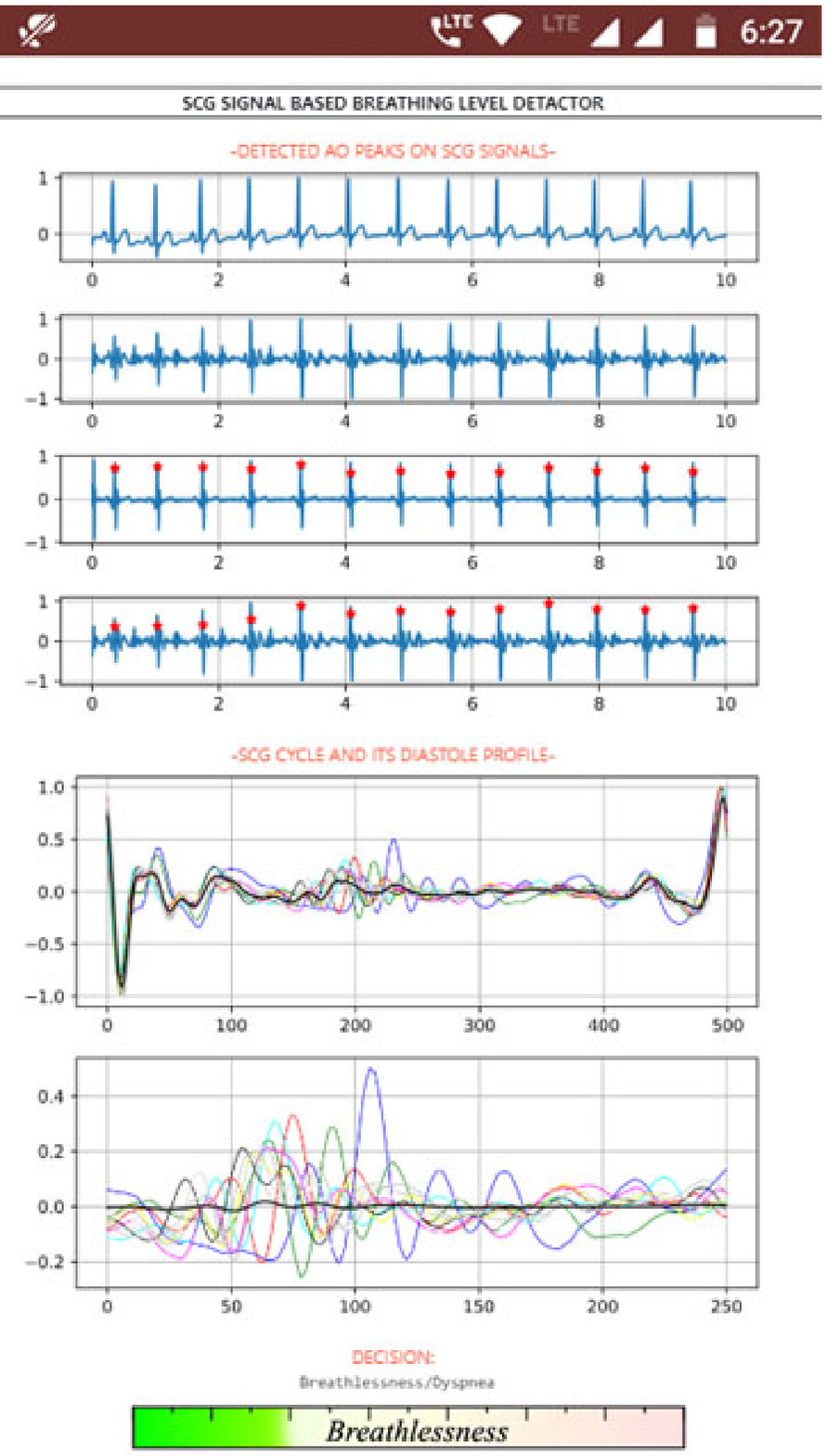}
\caption{Snapshots of our developed mobile application for breathing state detection.}
\label{fig:mobile_app}
\end{figure}

\begin{figure*}
\centering
\includegraphics[width=6.1cm, height=4.2cm]{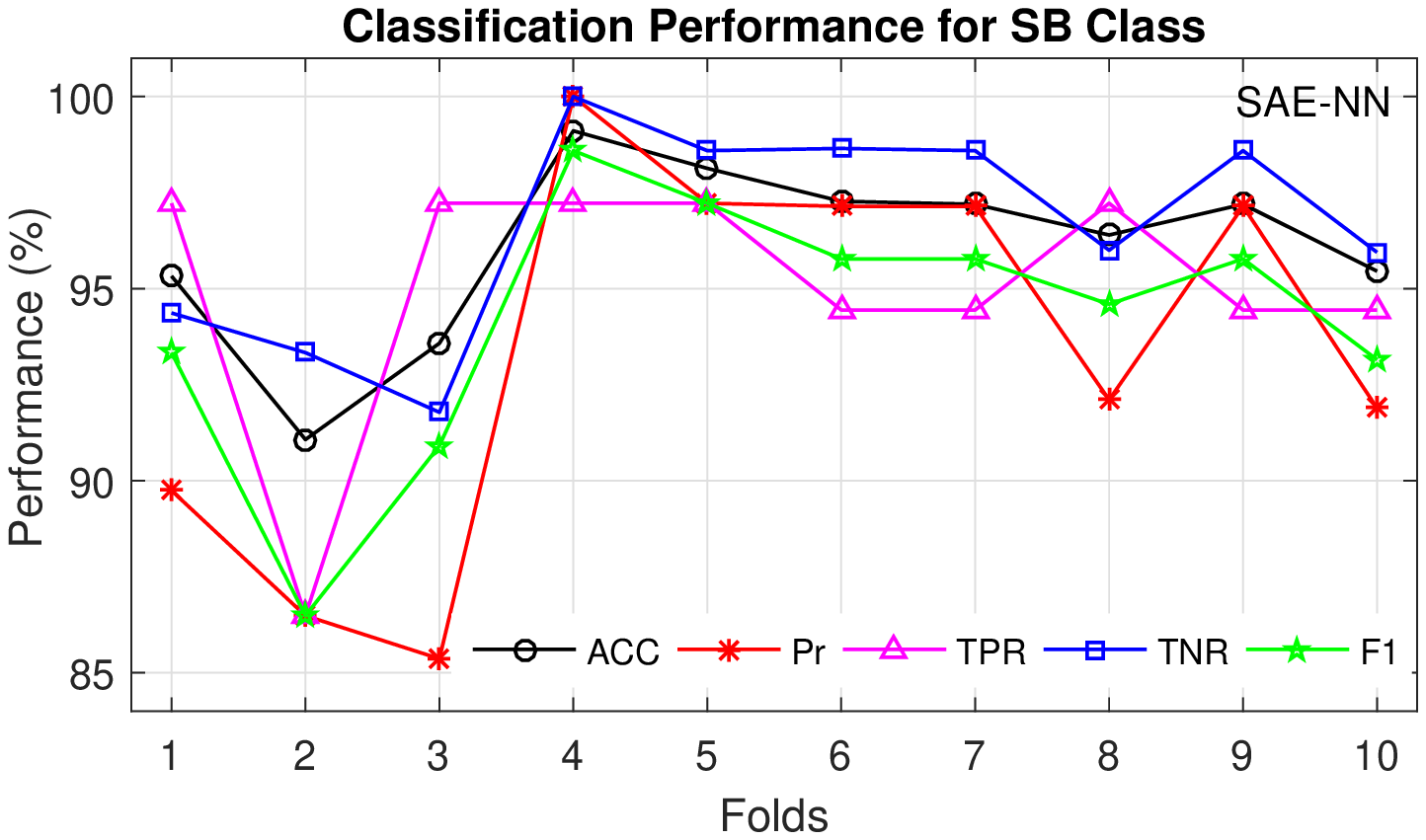}\hspace{-.2cm}
\includegraphics[width=6.1cm, height=4.2cm]{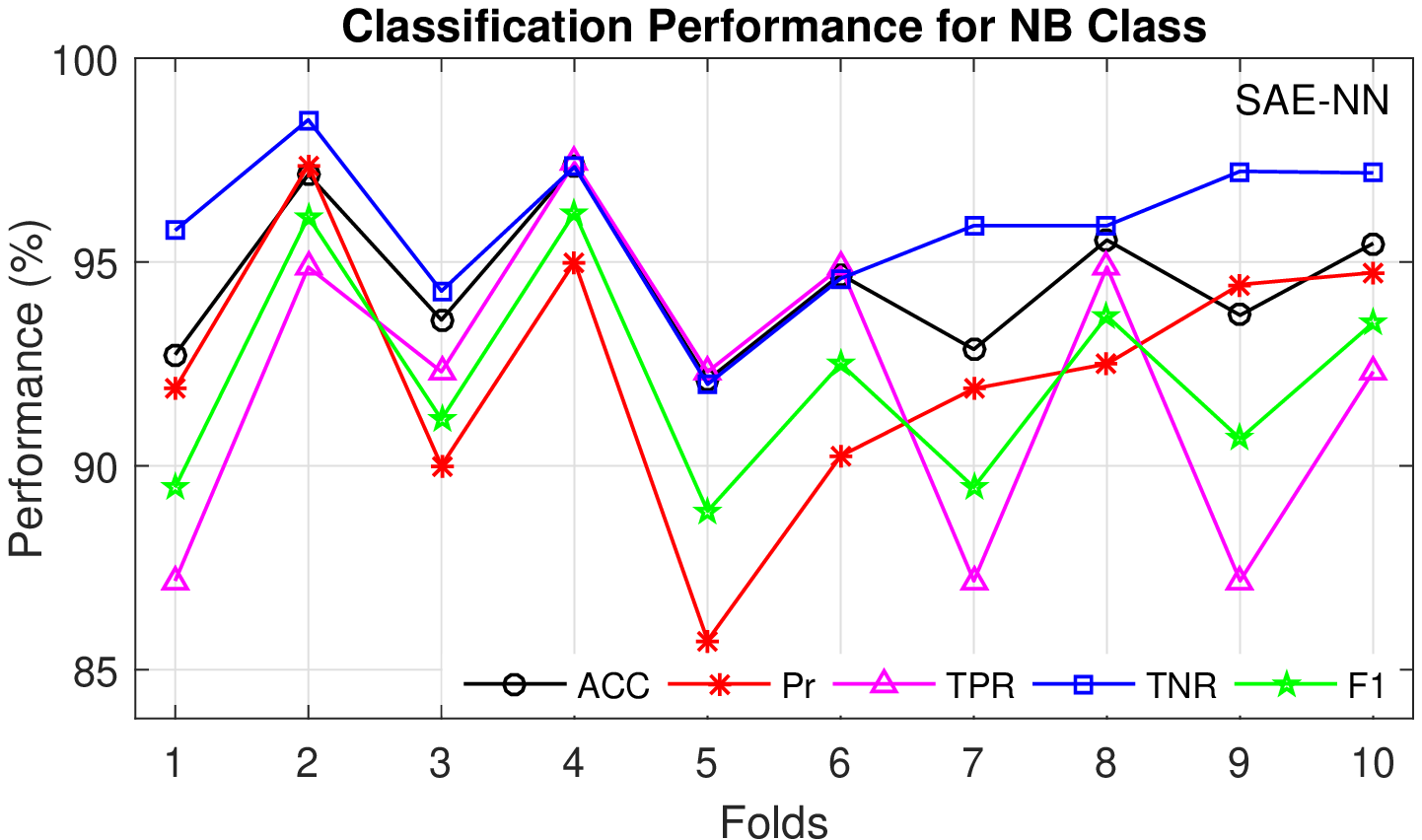}\hspace{-.2cm}
\includegraphics[width=6.1cm, height=4.2cm]{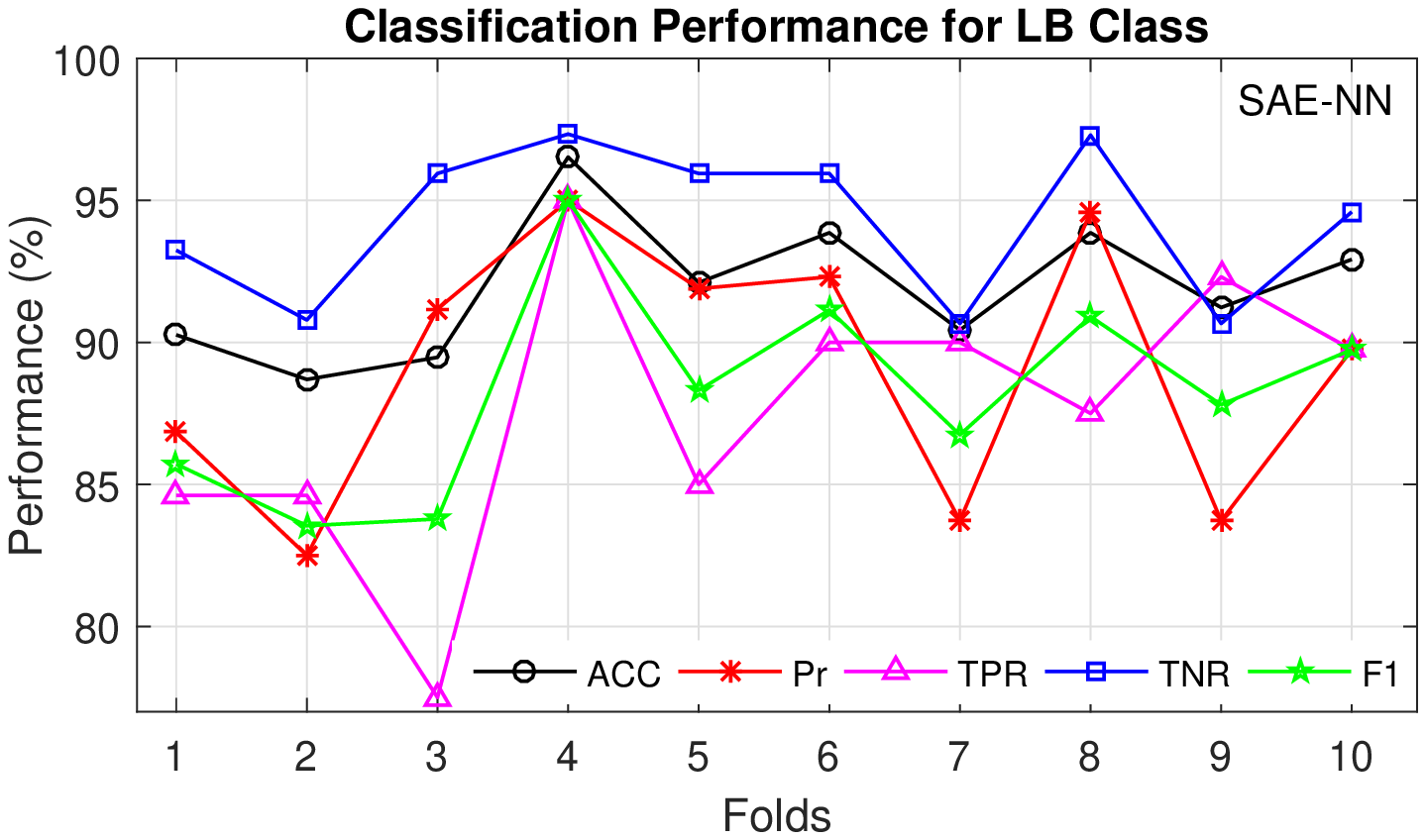}
\caption{Classification performance of the proposed method for identification of different respiratory-patterns.}
\label{fig:DNN_assessment}
\end{figure*}

Initially, OSP-based AO peak detection algorithm is employed to estimate AO peaks of an SCG signal. 
The detected AO peaks are used to identify SCG cycles. Subsequently, fifteen significant features are extracted on each of the SCG cycles. 
The final feature set is represented as:  
$f^{final}$=$\lbrace f^{HR}, f^{DHR}, f^{BEnr}, f^{BEnt}, f^{DBEnr}, f^{DBEnt}, f^{K},$ $ 
f^{IA}, f^{ACF}, f^{DEnt}, f^{DEnr}, f^{MSA}, f^{FMSA}, f^{BSEnt}, f^{BSC} \rbrace $
A NN architecture using SAE is proposed for classification, which can handle the feature engineering on its own. 
The parameters used in the proposed SAE are as follows:
number of neurons in hidden layers equal to 12 and 10, respectively. The values of $\lambda$ and $\beta$ are 0.001 and 4, respectively. The sparsity proportions for hidden layers are selected as 0.5 and 0.35, respectively.
The efficiency of the proposed approach is established based on standard quantitative statistical assessments. 
The performance measures used are recognition accuracy (ACC), precision (Pr), true positive rate (TPR), true negative rate (TNR) and F1-score. 
The aforementioned metrics can be computed using following expressions:
$\text{ACC}$ = $\frac{\text{TP}+\text{TN}}{\text{TP}+\text{TN}+\text{FP}+\text{FN}}$, 
$\text{Pr}$ = $\frac{\text{TP}}{\text{TP}+\text{FP}}$,
$\text{TPR}$ = $\frac{\text{TP}}{\text{TP}+\text{FN}}$,
$\text{TNR}$ = $\frac{\text{TN}}{\text{TN}+\text{FP}}$, and
$\text{F1-score}$ = $\frac{2 \times \text{Pr} \times \text{TPR}}{\text{Pr}+\text{TPR}}$.
The experimentation was performed with the data collected in three different sessions corresponding to SB, NB, and LB. 
Out of total 1147 observations, 1033 were used for training, and rest 114 were used for the testing process by employing 10-fold cross-validation methodology.
Table \ref{tab:DNNfold_Results} lists the foldwise recognition accuracies produced by the proposed method. The average overall recognition accuracy of 91.45\% is achieved.
Additionally, accuracies in identifying SB, NB, and LB breathing-patterns are determined.
The performance results are shown in Fig.~\ref{fig:DNN_assessment} for different folds.
It is observed that average accuracies achieved for classification of SB, NB and LB classes are in the range of 96--97\%, 94--95\%, and 91--92\%, respectively.
The average performance results for all these breathing classes are tabulated in Table~\ref{tab:Avg_Perf_DNN}. 
The average achieved precision rate from all three classes is 91.65\%.
The average TPR value obtained is in the range of 91--92\%. 
The averaged TNR and F1-score are approximately, 95.56\% and 91.53\%. 
All the performance results indicate the efficacy of the proposed method, and our proposed framework has a great potential in the domain of automatic identification of degree of breathing.
The snapshots of the developed mobile application are illustrated in Fig.~\ref{fig:mobile_app}. In this example, it is shown that an SCG signal of duration 10~s is recorded under breathlessness condition, and subsequently, it is successfully identified by the proposed framework.

\begin{table}
  \centering
  \caption{ Foldwise overall accuracies of the proposed SAE classifier}
   \newcommand{\m}{\hphantom{$-$}}
\newcommand{\cc}[1]{\multicolumn{1}{c}{#1}}
\renewcommand{\tabcolsep}{0.14pc} 
\renewcommand{\arraystretch}{0.92} 
\scriptsize
    \begin{tabular}{cccccccccccc}
    \toprule
    \textbf{Folds} & 1     & 2     & 3     & 4     & 5     & 6     & 7     & 8     & 9     & 10    & \textbf{Average} \\
    \midrule
    \textbf{ACC (\%)} & 89.47 & 88.70 & 88.70 & 96.52 & 91.30 & 93.04 & 90.43 & 93.04 & 91.23 & 92.11 & \textbf{91.45} \\
    \bottomrule
    \end{tabular}
  \label{tab:DNNfold_Results}
\end{table}%

\begin{table}[t!]
  \centering
  \caption{Average performance of the proposed SAE-based approach}
    \begin{tabular}{cccccc}
    \toprule
    Resp. Class & ACC   & Pr    & TPR   & TNR   & F1-score \\
    \midrule
    \multicolumn{1}{c}{SB} & \multicolumn{1}{c}{96.07} & \multicolumn{1}{c}{93.42} & \multicolumn{1}{c}{95.04} & \multicolumn{1}{c}{96.59} & \multicolumn{1}{c}{94.16} \\
    \multicolumn{1}{c}{NB} & \multicolumn{1}{c}{94.52} & \multicolumn{1}{c}{92.38} & \multicolumn{1}{c}{92.05} & \multicolumn{1}{c}{95.87} & \multicolumn{1}{c}{92.16} \\
    \multicolumn{1}{c}{LB} & \multicolumn{1}{c}{91.94} & \multicolumn{1}{c}{89.15} & \multicolumn{1}{c}{87.63} & \multicolumn{1}{c}{94.24} & \multicolumn{1}{c}{88.27} \\
    \cmidrule(){1-6}
    \multicolumn{1}{c}{\textbf{Average*}} & \multicolumn{1}{c}{\textbf{94.17}} & \multicolumn{1}{c}{\textbf{91.65}} & \multicolumn{1}{c}{\textbf{91.57}} & \multicolumn{1}{c}{\textbf{95.56}} & \multicolumn{1}{c}{\textbf{91.53}} \\
    \bottomrule
    \vspace{-0.4cm}
    \end{tabular}%

\scriptsize      
  \vspace{0.3cm}
  Note that ACC, Pr, TPR, TNR and F1-scores are given in \%. 
  
  *Average performance indexes for all respiratory levels.  
  \label{tab:Avg_Perf_DNN}%
\end{table}%

\begin{figure*}[!h]
\centering
\includegraphics[width=6cm]{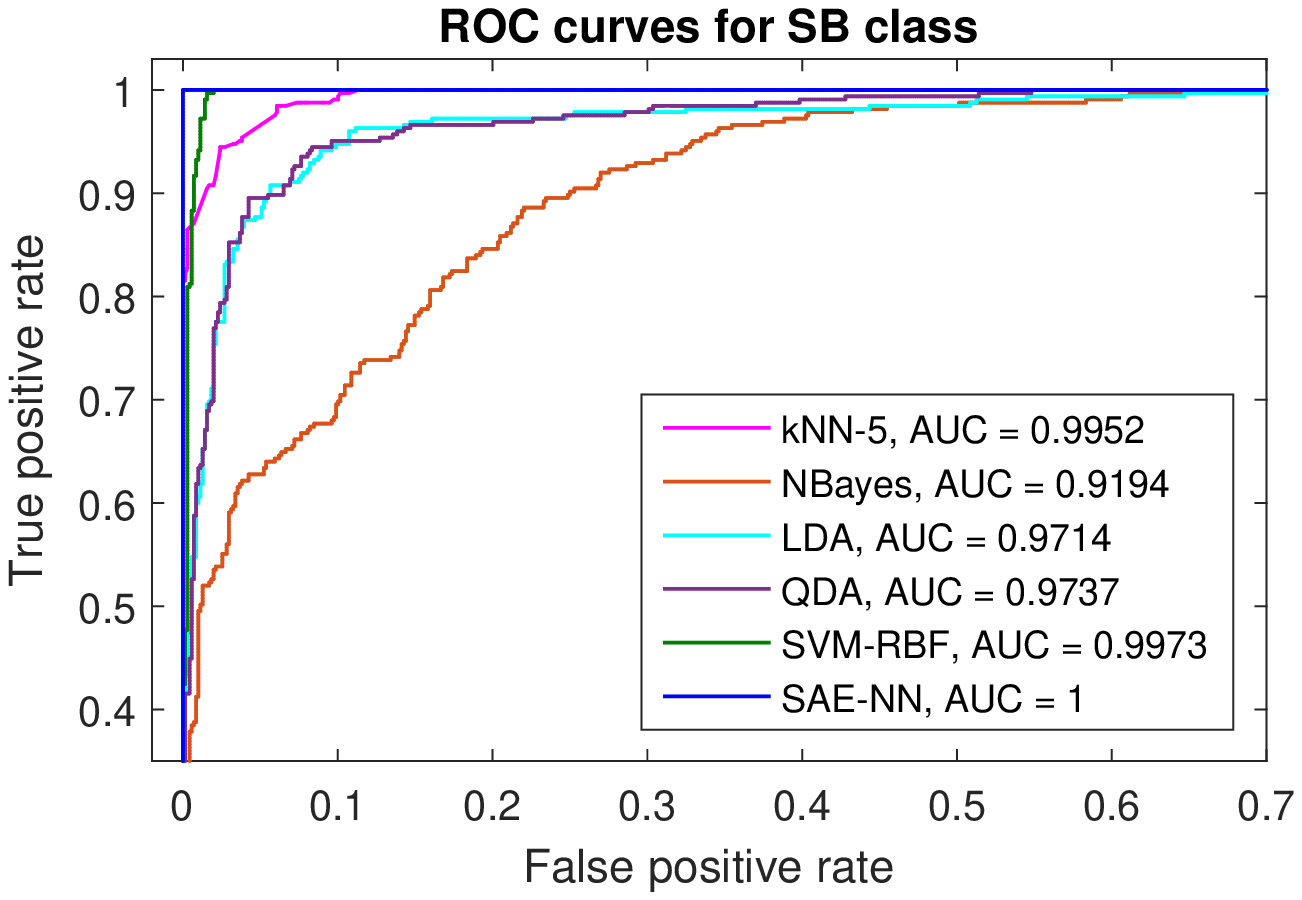}\hspace{-0.1cm}
\includegraphics[width=6cm]{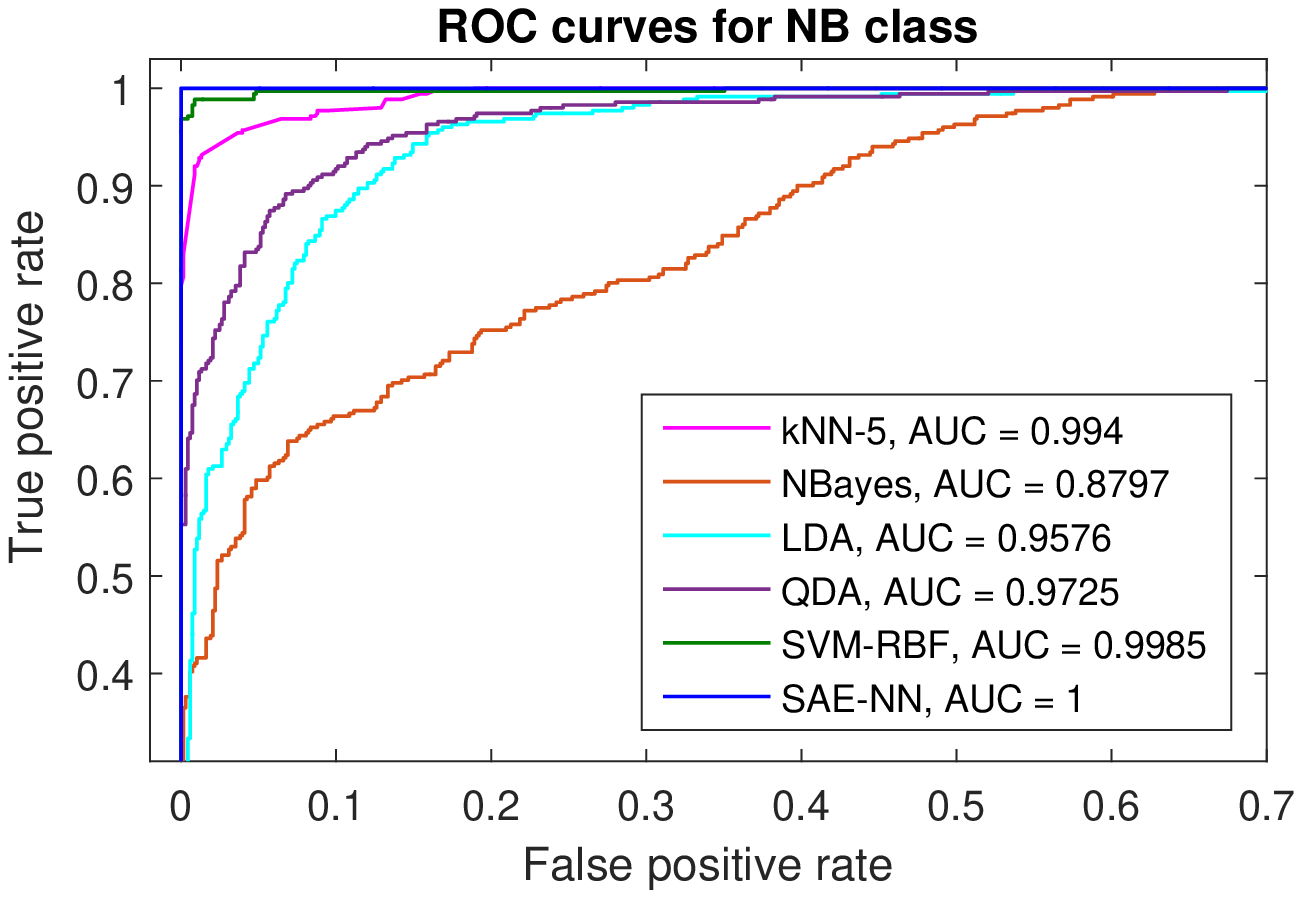}\hspace{-0.1cm}
\includegraphics[width=6cm]{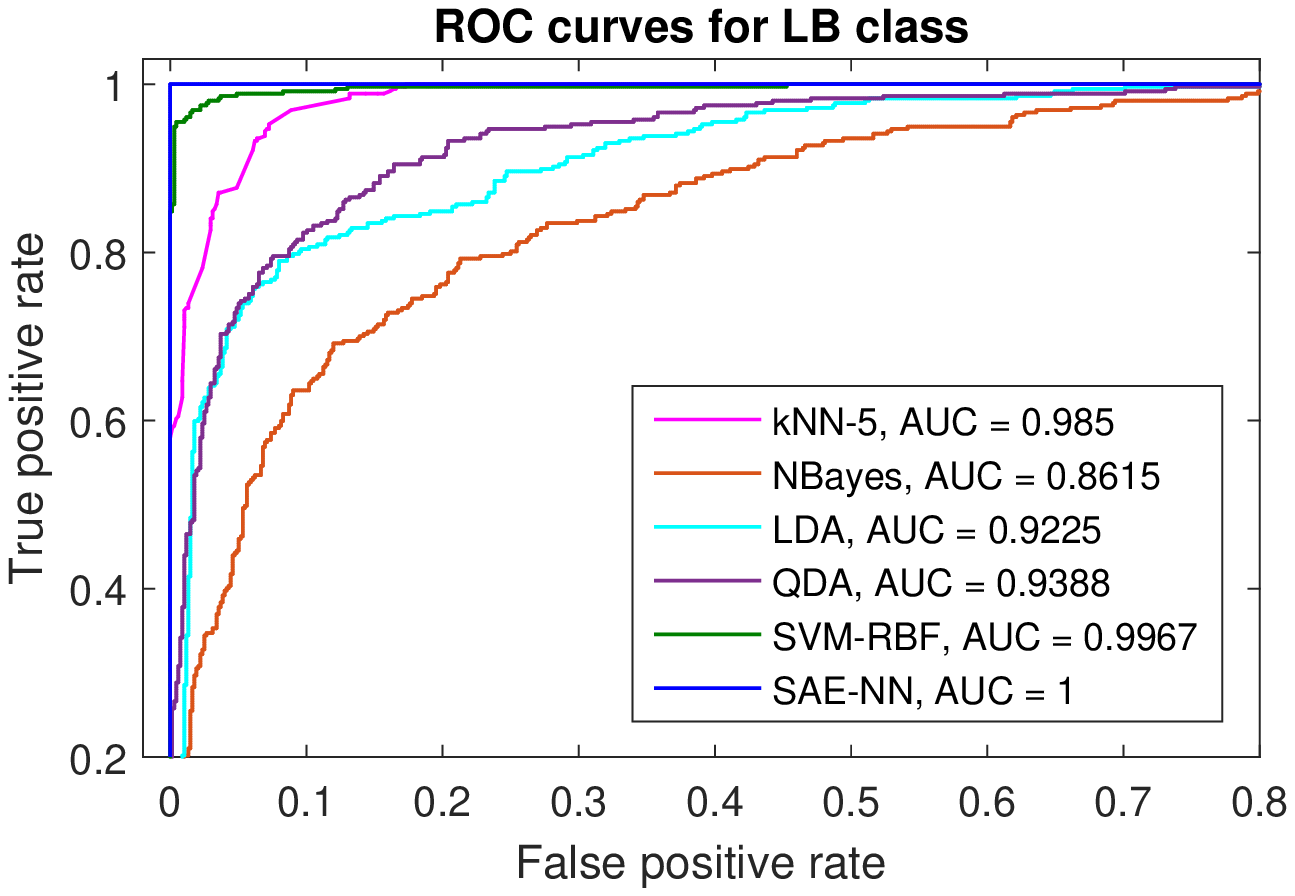}
\caption{ROC curves of different classifiers. Note that AUC represents area under the ROC curve and false positive rate is computed as: $\text{FPR} = 1 - \text{TNR}$.}
\label{roc}
\end{figure*}

\subsection{Performance Comparison}
The proposed technique is compared with other conventional classifiers namely, SVM (with different kernels such as RBF, linear and polynomial), kNN (with different number of neighbours), naive Bayes, ensemble classifiers, linear discriminant analysis (LDA), and quadratic discriminant analysis (QDA) based classifiers. 
 We deployed a statistical based feature analysis technique called multi-variate feature analysis (MANOVA) \cite{hastie2005} technique to select dominant features for comparison of different classifiers. 
All the fifteen features are accepted as all of them show more than 5\% level of significance ($p \leq 0.05$). 
The performance comparison of our method with different classifiers is shown in Table \ref{tab:comparison}. 
The results clearly show that the proposed method outperforms other conventional methods.
Also, ROC curves for all these classifiers are shown in Fig. \ref{roc}.

\begin{table}[t!]
  \centering
  \caption{Performance comparison in terms of overall accuracy (\%) obtained from 10-folds experimentation}
  \newcommand{\m}{\hphantom{$-$}}
\newcommand{\cc}[1]{\multicolumn{1}{c}{#1}}
\renewcommand{\tabcolsep}{0.19pc} 
\renewcommand{\arraystretch}{0.92} 
    \begin{tabular}{cccccccccc}
    \toprule
   Proposed & \multicolumn{3}{c}{SVM}   & \multicolumn{3}{c}{kNN}  & Naive & \multirow{2}[4]{*}{LDA} & \multirow{2}[4]{*}{QDA} \\
    \cmidrule(r){2-4}   \cmidrule(l){5-7} 
   SAE-NN       & RBF   & Linear & Poly3 & K=5 & K=10 & K=100 &  Bayes     &       &  \\
          \midrule
    \textbf{91.45} & 90.50 & 86.84 & 89.19 & 89.45 & 87.53 & 77.52 & 71.32 & 85.10 & 85.01 \\
    \bottomrule
    \end{tabular}%
    
    \vspace{0.2cm}
        \begin{tabular}{cccccc}
    \toprule
   \multicolumn{3}{c}{Random forest}   & \multicolumn{3}{c}{AdaBoost (Ns = 50)}   \\
    \cmidrule(r){1-3}   \cmidrule(l){4-6} 
   Nl = 30   & Nl = 100 & Nl = 200          & Nl = 100   & Nl = 200 & Nl = 300 \\
          \midrule
     89.1 & 90.8 & 90.1 & 90 & 90.5 & 90.3  \\
    \bottomrule
    \end{tabular}%
    
    \vspace{0.1cm}
    Nl and Ns denote number of learners and number of splits, respectively.
  \label{tab:comparison}%
\end{table}

\section{Conclusion}
In this work, an SCG based breathing-state detector is developed for m-healthcare applications.  
The proposed method can accurately identify the degree-of-breathings, such as breathlessness, normal breathing, and long and labored breathing conditions.
The concurrent ECG signal is also used to extract the SCG cycles using OSP-based scheme, and each of the cycles is used for feature extraction.
A set of robust and simple features is extracted from the SCG signal, which conveys the information of hemodynamic changes and physiological movement of lung and heart muscles due to varying breathing-rates.
For classification of different breathing patterns, SAE-based NN architecture is proposed.
The performance of our method is evaluated on 1147 SCG cycles in different breathing scenarios.  
The quantitative-assessment results clearly show that the proposed method may be deployed for many day-to-day life and clinical applications. 
In this study, the performance of the proposed method was evaluated for data collected at rest in controlled experimental conditions, and hence, this research could also be extended for different physiological modulations and pathological conditions in ambulatory and in-house environments.
The limitation of the proposed method is that it uses two cardiac modalities, SCG and ECG, which needs more than one sensing device. The overall process slightly reduces the comfortability of the user. The use of ECG signals could have been avoided by developing a robust standalone AO instant detection framework for the SCG.
One major finding of this research work is that the SCG signal can be used not only for cardiac health measurement, but also for the assessment of respiration-rates and lung fitness.

\bibliographystyle{IEEEtran}
\bibliography{tc_References}
\end{document}